\documentclass[12pt]{article}

\usepackage{color,graphics,latexsym,citesort}

\newcommand{\SF}{\ensuremath{\mathcal{A}}}
\newcommand{\SH}{\ensuremath{\mathcal{C}}}

\newcommand{\tr}{\ensuremath{\mathrm{tr}\,}}
\newcommand{\str}{\ensuremath{\mathrm{str}\,}}
\newcommand{\Int}[1]{\ensuremath{\int \!\! d^D \! #1 \,}}
\newcommand{\volume}[1]{\ensuremath{d^D \! #1 \,}}

\newcommand{\one}{\hbox{1\kern-.8mm l}}
\newcommand{\pder}[2]{\ensuremath{\frac{\partial #1}{\partial #2}}}
\newcommand{\fder}[2]{\ensuremath{\frac{\delta #1}{\delta #2}}}

\newcommand{\PhysRev}[4]{\emph{Phys.\ Rev.\ }\textbf{#1 #2} (#3) #4}
\newcommand{\PhysRep}[4]{\emph{Phys.\ Rep.\ }\textbf{#1 #2} (#3) #4}
\newcommand{\IntJModPhys}[4]{\emph{Int.\ J.\ Mod.\ Phys.\ }\textbf{#1 #2} (#3) #4}
\newcommand{\NuclPhys}[4]{\emph{Nucl.\ Phys.\ }\textbf{#1 #2} (#3) #4}

\begin{document}
\begin{center}
\Large
\textbf{The Gauge Invariant ERG}
\normalsize

\vspace{2ex}
$\mbox{O.J.~Rosten}^{\dagger \;}$\footnote{Based on a talk given by this author at Quarks-2004.}, $\mbox{T.R.~Morris}^{\ddagger}$ and $\mbox{S.~Arnone}^{\S}$
\vspace{2ex}

$\left.^{\dagger}\right.\left.^{\ddagger}\right.$ \emph{School of Physics \& Astronomy, University of Southampton, Highfield, Southampton S017 1BJ, U.K.}

$\left.^{\S}\right.$ \emph{Dipartimento di Fisica,
Universit\`a di Roma ``La Sapienza'', \\
P.le Aldo Moro, 2 - 00185 Roma - Italy}
\end{center}

\begin{flushleft}
Emails: $\left.^{\dagger}\right.$O.J.Rosten@soton.ac.uk, $\left.^{\ddagger}\right.$T.R.Morris@soton.ac.uk \\ 
and $\left.^{\S}\right.$Stefano.Arnone@roma1.infn.it.
\end{flushleft}

\begin{abstract}

We sketch the construction of a gauge invariant Exact Renormalization Group (ERG).
Starting from Polchinski's equation, the emphasis is on how a series
of ideas have combined to yield the gauge invariant formalism. 

A novel symmetry of the ERG allows the flow equation to be modified, in
such a way that it is  suitable for the computation of the (universal)
two-loop $\beta$-function. This computation has now been performed, within
the framework of the ERG and, as such, in a \emph{manifestly gauge invariant} way
for the very first time.

\end{abstract}

\section{Introduction} \label{sec:intro}

Two of the most powerful ideas in quantum field theory (QFT) are those of Wilson's renormalization
group (RG)~\cite{Wilson} and gauge symmetries. In this report, we describe a framework
which incorporates both and, as such, may provide a powerful tool for the investigation of
a host of problems within gauge theory. It is particularly noteworthy that, within this
set-up, gauge invariance is manifest, allowing calculations to be done, without 
fixing the gauge~\cite{YM-1-loop}.

We will concern ourselves with exact formulations of the renormalization group~\cite{Wilson,TRM-elements,Pol}
which provide an equation describing how the effective action at some cutoff scale varies with this scale. 
To build towards a gauge invariant ERG, we begin by looking at scalar field theory, in section~\ref{sec:Pol}.
Starting from Polchinski's equation~\cite{Pol}, we generalise it to bring it to a form suitable for
the construction of a gauge invariant ERG.

Applying the methods of the ERG to gauge theories was long beset by the difficulty of introducing a regulator
which suppresses modes above $\Lambda$, whilst maintaining gauge invariance. In section~\ref{sec:reg},
we begin by describing a regulator for Yang-Mills theory, suitable for use within the ERG~\cite{SU(N|N)}.
Next we discuss a novel symmetry of the regularization scheme, which proves crucial in the generalisation
of the ERG to a form suitable for computing the two-loop $\beta$-function.

The gauge invariant ERG of ref.~\cite{YM-1-loop} is introduced in section~\ref{sec:GI-ERG}. After
outlining its construction, we discuss the criterion for obtaining the universal value
for $\beta_1$ and $\beta_2$. This leads us to a modified form of the flow equation.
Finally, we outline how $\beta$-function coefficients can be extracted.

\section{Generalising Polchinski's Equation} \label{sec:Pol}

\begin{figure}[h]
\begin{center}
\input{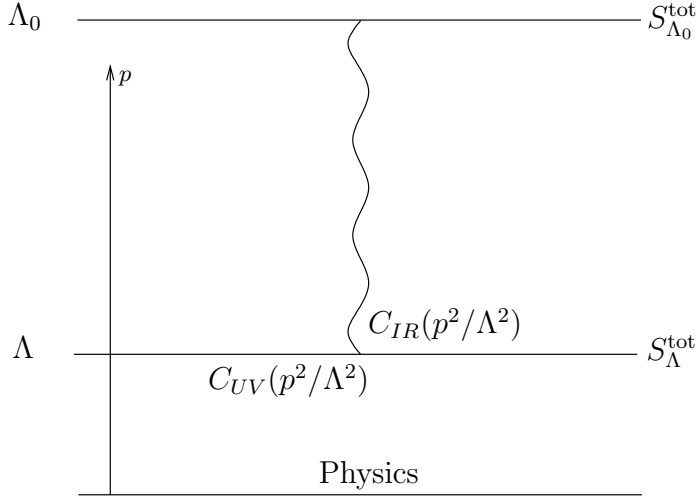}
\caption{Integrating out modes.}
\label{fig:ERG}
\end{center}
\end{figure}

The key elements of the ERG are depicted in figure~\ref{fig:ERG}~\cite{TRM-elements}. Starting with the bare 
theory---for the scalar field $\varphi$---defined at the scale $\Lambda_0$, and 
parametrized by the bare action $S_{\Lambda_0}^\mathrm{tot}$,
we integrate out degrees of freedom, to some intermediate scale, $\Lambda$. This scale separates
the high energy modes from the low energy modes and can be viewed in one of two ways: for the
high energy modes, it acts as an infra-red (IR) regulator whereas, for the low energy modes, it
acts as an ultra-violet (UV) regulator. In recognition of this interpretation,
we introduce two cutoff functions,
$C_{\mathrm{IR}}(p^2/\Lambda^2)$ and $C_{\mathrm{UV}}(p^2/\Lambda^2)$, each of which cuts off modes
in the region indicated by the subscript. We leave these cutoff functions general, demanding
only that
\[
C_{\mathrm{IR}}(p^2/\Lambda^2) + C_{\mathrm{UV}}(p^2/\Lambda^2) = 1
\]
and that
\begin{eqnarray*}
C_{\mathrm{UV}}(0) & = & 1;\\
\lim_{z\rightarrow\infty} C_{\mathrm{UV}}(z) &\rightarrow& 0 \ \mbox{(fast  enough).} 
\end{eqnarray*}
In the high energy and low energy regions, we now modify the propagators by multiplying
by the cutoff functions:
\[
1/p^2 \rightarrow 
\left\{
\begin{array}{llll}
\Delta_{UV} & = & \frac{C_{\mathrm{UV}}(p^2/\Lambda^2)}{p^2} & \mbox{low energy} \\
\Delta_{IR} & = & \frac{C_{\mathrm{IR}}(p^2/\Lambda^2)}{p^2} & \mbox{high energy}.
\end{array}
\right.
\]
By performing the integral over high energy modes, we can obtain Polchinski's equation~\cite{Pol}
for the flow of the interaction part of the Wilsonian effective action, $S_\Lambda^{\mathrm{int}}$:
\begin{equation}
\pder{}{\Lambda} S_{\Lambda}^{\mathrm{int}}[\varphi] = \frac{1}{2} \fder{S_\Lambda^{\mathrm{int}}}{\varphi} \cdot \pder{\Delta_{UV}}{\Lambda} \cdot
\fder{S_\Lambda^{\mathrm{int}}}{\varphi} 
- \frac{1}{2} \fder{}{\varphi} \cdot \pder{\Delta_{UV}}{\Lambda} \cdot \fder{S_\Lambda^{\mathrm{int}}}{\varphi},
\label{eq:Pol}
\end{equation}
where, for the functions $f(x)$ and $g(y)$ and a momentum space kernel $W(p^2/\Lambda^2)$,
\begin{equation}
f \cdot W \cdot g = \int \!\! \int d^Dx \, d^Dy \, f(x) W_{xy} \, g(y),
\label{eq:f.W.g}
\end{equation}
with 
\begin{equation}
W_{xy} = \Int{p} W(p^2/\Lambda^2)e^{ip.(x-y)}.
\label{eq:PosSpaceKernel}
\end{equation}

There are three crucial properties of the Polchinski equation:
\begin{enumerate}
	\item the flow in $S$ corresponds to integrating out higher energy modes;
	
	\item the partition function is left invariant under the flow;

	\item it is defined non-perturbatively.
\end{enumerate}
The point that will ultimately enable us to construct a gauge invariant flow equation is
that other flow equations have these properties; indeed, from now on, we take these properties
(in particular the first two) to define what is meant by a flow equation~\cite{YM-1-loop,TRM+JL}.

The first step in the necessary generalisation of Polchinski's equation is trivial: we
change variables in order to compute the flow of the total Wilsonian effective action,
$S$. To this end, our regularised kinetic term is
\[
\hat{S} = \frac{1}{2} \varphi \cdot \Delta_{UV}^{-1} \cdot \varphi,
\]
and so $S = \hat{S} + S_\Lambda^{\mathrm{int}}$. We will call $\hat{S}$ the \emph{seed action}.
Introducing the new variable
\[
\Sigma = S - 2 \hat{S},
\]
and defining $\dot{X} = -\Lambda \partial_\Lambda X$,
we can rewrite Polchinski's equation,  up to a discarded vacuum energy term, as
\[
\dot{S} = \frac{1}{2} \fder{S}{\varphi} \cdot \dot{\Delta} \cdot
\fder{\Sigma}{\varphi} 
- \frac{1}{2} \fder{}{\varphi} \cdot \dot{\Delta} \cdot \fder{\Sigma}{\varphi}.
\]

Now, however, we perform a non-trivial step: we allow $\hat{S}$ to become very
general~\cite{YM-1-loop} but keep $S$ as the total Wilsonian effective action. In other words,
the seed action is now defined independently of the total Wilsonian effective action.
The question now is whether our flow equation is still valid. The partition function
remains invariant under the flow:
\[
\Lambda \pder{}{\Lambda} e^{-S} = -\frac{1}{2} \fder{}{\varphi} \cdot \dot{\Delta} \cdot \left(\fder{\Sigma}{\varphi} e^{-S}
\right).
\]
Moreover, the flow still corresponds to integrating out:
\begin{quotation}
We
can ensure that the flow equation is regularized, so all momentum
integrals are bounded.
$\Lambda$ is the UV cutoff: momenta larger than some scale $q$ must vanish in the limit $q/\Lambda \rightarrow \infty$.
As $\Lambda \rightarrow 0$ all remaining contributions from any non-vanishing momentum scale disappear.
\emph{But} the physics is invariant under the flow.
Contributions from a given momentum scale must be encoded in the effective action: we have integrated out!
\end{quotation}
As our flow equation satisfies the requisite requirements, it is perfectly valid. This generalisation is one of
the necessary changes required for Yang-Mills theory~\cite{YM-1-loop}. Moreover, the great freedom
in the seed action actually allows us to design it in order to simplify calculations!

\section{Regularization for Yang-Mills Theory} \label{sec:reg}

\subsection{The Construction}

The final ingredient for our gauge invariant ERG is a regulator for Yang-Mills theory, based on a
cutoff. To ensure compatibility with gauge invariance we cast equation~\ref{eq:f.W.g} in gauge
invariant form, replacing $f$ and $g$ with any two matrix representations, $u$ and $v$:
\begin{eqnarray}
\lefteqn{
u\{W\} v = 
\sum_{m,n=0}^{\infty} \Int{x}\volume{y}\volume{x_1}\cdots\volume{x_n}\volume{y_1}\cdots\volume{y_m} 
} \nonumber \\
& & W_{\mu_1\cdots\mu_n,\nu_1\cdots\nu_m}(x_1,\ldots,x_n;y_1,\ldots,y_m;x,y) \nonumber \\
& & \tr \left[u(x)A_{\mu_1}(x_1)\cdots A_{\mu_n}(x_n)v(y)A_{\nu_1}(y_1)\cdots A_{\nu_m}(y_m) \right],
\label{eq:Wine}
\end{eqnarray}
where $\{W\}$, the `wine', is the gauge covariantization of the kernel $W_{xy}$ and equation~\ref{eq:Wine}
defines what we mean by the wine vertices.
Note that the case $m=n=0$ is just the original kernel $W_{xy}$, which we call a zero-point wine. 
We refer to the remaining higher-point wine vertices as `decorations' of the zero-point wine.

We can now write the
Yang-Mills kinetic term as
\[
\frac{1}{2} F_{\mu \nu} \{c^{-1}\} F_{\mu \nu},
\]
where $c$ is a (UV) cutoff function and $\{c^{-1}\}$ the covariantization of its inverse.

However, this turns out to be insufficient to regularise the theory. The solution
we choose is to embed our $SU(N)$ gauge theory in an $SU(N|N)$ gauge theory, which has sufficiently
improved UV properties to ensure finiteness~\cite{SU(N|N)}. 

Taking the $SU(N|N)$ gauge field to be $\SF_\mu$, we embed our
physical $SU(N)$ field $A^1_\mu$ using the defining representation:
\begin{equation}
\SF_\mu = \SF_\mu^0 \one + \left(
	\begin{array}{cc}
		A_\mu^1 & B_\mu \\
		\bar{B}_\mu & A_\mu^2
	\end{array} \right).
\label{eq:SuperGaugeField}
\end{equation}
The supermatrix possesses bosonic block diagonal fields, $A^i$, fermionic off-diagonal fields $B, \ \bar{B}$
and the central term $\SF_0 \one$. As required by $SU(N|N)$, the superfield is supertraceless:
$\str \SF_\mu = \tr A^1_\mu - \tr A^2_\mu = 0$. The covariant derivative is simply
\[
\nabla_\mu = \partial_\mu - i \SF_\mu,
\]
where, in readiness for section~\ref{sec:GI-ERG}, we have absorbed the coupling constant into the
gauge field. The field strength is $\mathcal{F}_{\mu \nu} = i[\nabla_\mu, \nabla_\nu]$ and
the regularised kinetic term just
\[
\frac{1}{2} \mathcal{F}_{\mu \nu} \{c^{-1}\} \mathcal{F}_{\mu \nu}.
\]
Finally, we introduce the superscalar field, $\SH$, to spontaneously break the $SU(N|N)$ symmetry in the
fermionic directions. The fermionic fields acquire a mass, of the order the cutoff, and act
as a set of Pauli-Villars (PV) fields. The problem of overlapping divergences, typical of
PV regularisation, never arises as the covariant cutoff regularisation applies 
to all fields.

Equation~\ref{eq:Wine} must now be modified to take account of the change from gauge invariance to supergauge 
invariance and the presence of $\SH$s. For the former, we simply substitute $\SF$s for $A$s and
replace the trace in equation~\ref{eq:Wine}
with a supertrace; for the latter, we now allow (up to some restrictions) decoration with both $\SF$s and $\SH$s~\cite{YM-1-loop}.

Being supergauge invariant, the action has an expansion in terms of supertraces and 
products of supertraces. Suppressing position arguments and Lorentz indices we have:
\begin{eqnarray*}
S & = & \sum_{n=1}^{\infty} \frac{1}{s_n} S^{X_1\cdots X_n}\str \left( X_1\cdots X_n \right) \\
  & + & \frac{1}{2!}\sum_{m,n=1}^{\infty}\frac{1}{s_n s_m} S^{X_1\cdots X_n, Y_1\cdots Y_m} \str \left( X_1\cdots X_n \right)
\str \left( Y_1\cdots Y_m \right) + \cdots
\end{eqnarray*}
where the $X_i$ are $\SF_\mu$ or $\SH$. Only one cyclic ordering of each list appears in the sum;
if a list is invariant under some nontrivial cyclic permutation, then $s_n$ ($s_m$) is the order of
the subgroup.

\subsection{No-$\SF^0$ Symmetry}

The structure of $SU(N|N)$ leads to a symmetry concerning the field
$\SF^0$.
The generator associated with $\SF^0$ is $\one$ and, since this commutes with everything, $\SF^0$ does
not have a kinetic term. Were $\SF^0$ to appear anywhere else in the action, it would
act as a Lagrange multiplier, enforcing a non-linear constraint on the theory. We thus
demand that the action is independent of $\SF^0$.
The theory is now invariant under a local `no-$\SF^0$' symmetry: $\delta \SF^0_\mu(x) = \lambda_\mu(x)$.
A full understanding of this symmetry has proven vital in performing calculations beyond one-loop~\cite{YM-2-loop}.

\section{A Gauge Invariant ERG} \label{sec:GI-ERG}

\subsection{Construction}

We start with an Abelian theory and make the simplest possible generalisation of the Polchinski equation:
\[
\dot{S} = \frac{1}{2} \fder{S}{A_\mu} \cdot \dot{\Delta} \cdot
\fder{\Sigma}{A_\mu} 
- \frac{1}{2} \fder{}{A_\mu} \cdot \dot{\Delta} \cdot \fder{\Sigma}{A_\mu}.
\]
This still has all the features of a flow equation that we want but 
we have the added benefit of not having fixed the gauge.

When we generalise to non-Abelian gauge theory, the exact preservation of the form of gauge transformations
\[
\delta A_\mu = [D_\mu, \omega]
\]
implies that $D_\mu = \partial_\mu - igA_\mu$ cannot renormalize~\cite{TRM1}. We make use of
this by changing variables: 
$A_\mu \rightarrow A_\mu /g, \
\hat{S} \rightarrow \hat{S} / g^2.$
The first change means that $D_\mu = \partial_\mu - iA_\mu$ giving the nice result that the gauge field
does not renormalize. Implementing both of our changes modifies the flow equation. One part
takes the same form as before, but there is a second term which is not manifestly gauge invariant.
Nonetheless, we can simply drop this term: the flow equation still leaves the partition function
invariant and still corresponds to integrating out~\cite{YM-1-loop}.
Our Abelian flow equation becomes:
\[
\dot{S} = \frac{1}{2} \fder{S}{A_\mu} \cdot \dot{\Delta} \cdot
\fder{\Sigma_g}{A_\mu} 
- \frac{1}{2} \fder{}{A_\mu} \cdot \dot{\Delta} \cdot \fder{\Sigma_g}{A_\mu}
\]
where $\Sigma_g = g^2S - 2\hat{S}$.

It is now straightforward to generalise to the non-Abelian case: we must simply
covariantize the cutoff and incorporate the $SU(N|N)$ regularization.
The fully regularized non-Abelian flow equation is:
\begin{equation}
\dot{S} = a_0[S,\Sigma_g] - a_1[\Sigma_g]
\label{eq:FlowEq}
\end{equation}
where
\[
a_0[S,\Sigma_g] = \frac{1}{2} \fder{S}{\SF_\mu} \{\dot{\Delta}^{\SF \SF}\} \fder{\Sigma_g}{\SF_\mu}
	+  \frac{1}{2} \fder{S}{\SH} \{\dot{\Delta}^{\SH \SH}\} \fder{\Sigma_g}{\SH}
\]
and
\[
a_1[\Sigma_g] = \frac{1}{2} \fder{}{\SF_\mu} \{\dot{\Delta}^{\SF \SF}\} \fder{\Sigma_g}{\SF_\mu}
	+  \frac{1}{2} \fder{}{\SH} \{\dot{\Delta}^{\SH \SH}\} \fder{\Sigma_g}{\SH}.
\]
The renormalized coupling, $g$, of the $SU(N)$ gauge field $A^1_\mu$ is defined through
the renormalization condition:
\begin{equation}
S[\SF = A^1, \SH = \bar{\SH}] = \frac{1}{2g^2} \mathrm{tr} \Int{x} (F_{\mu \nu}^1)^2 + \cdots
\label{eq:renorm}
\end{equation}
where $\bar{\SH}$ is the \emph{vev} of the field $\SH$ and the ellipsis denote higher dimension operators.

To complete the set up, we work in the broken phase and 
determine the zero-point wines, $\dot{\Delta}^{XX}$. The fields $X$ can be any of the broken phase fields. 
The current flow equation cannot distinguish between $A^1$ and $A^2$ and so 
we treat them together, within the block-diagonal field $A$.
The first step is to expand the flow equation for small coupling. We have the following expansions:
\begin{eqnarray}
S & = & 1/g^2 S_0 + S_1 + g^2 S_2 + \cdots \nonumber\\
\Lambda \pder{g}{\Lambda} & = &\beta_1 g^3 + \beta_2 g^5 + \cdots \label{eq:beta}
\end{eqnarray}
and, introducing $\Sigma_n = S_n -2 \hat{S}_n$, the weak coupling flow equations:
\begin{eqnarray}
\dot{S_0} & = & a_0[S_0,\Sigma_0] \label{eq:classicalFlow}\\
\dot{S_1} & = & -2 \beta_1 S_0 + a_0[S_1,\Sigma_0] + a_0[S_0,\Sigma_1] - a_1[\Sigma_1] \nonumber\\
\vdots & & \nonumber
\end{eqnarray}

Now we  determine the two-point tree level seed action vertices by using
Lorentz invariance, dimensions and the renormalization condition.
In this way, we determine that~\cite{YM-1-loop}
\[
\hat{S}^{\ AA}_{0 \mu \nu}(p) = 2 \Box_{\mu \nu}(p) /c_p,
\]
where  the renormalization condition demands that $c(0)=1$.

Using equation~\ref{eq:classicalFlow} to compute the flow of the two-point tree level Wilsonian effective
action vertex $S^{\ AA}_{0 \mu \nu}(p)$, we obtain an equation relating this to the corresponding
seed action vertex and $\dot{\Delta}^{AA}$. 
By choosing to set $S^{\ AA}_{0 \mu \nu}(p) = \hat{S}^{\ AA}_{0 \mu \nu}(p)$, we find that 
$\dot{\Delta}^{AA} = c'_p/\Lambda^2$. In a similar fashion, we can compute the remaining
zero-point wines.

This is the set-up used in ref.~\cite{YM-1-loop} to compute the
one-loop $\beta$-function, without gauge fixing.

\subsection{Universality of $\beta_1$ and $\beta_2$}

Given the coupling $g$, within our renormalization scheme, we can perturbatively relate this to that
of another renormalization scheme, $\tilde{g}$:
\[
1/\tilde{g}^2 = 1/g^2 + \gamma + \mathcal{O}(g^2)
\]
where $\gamma$ is a matching coefficient. From equation~\ref{eq:beta} we have:
\[
\tilde{\beta}_1 + \tilde{\beta}_2 g^2 = \beta_1 + \beta_2g^2 - \dot{\gamma} + \mathcal{O}(g^4).
\]
Agreement between the first two $\beta$-function coefficients is guaranteed, so long
as $\gamma$ does not run at either the tree or one-loop level.

Incorporation of PV fields into the ERG in fact generically introduces
tree-level running. However, this can be removed by suitable choice of the
seed action~\cite{YM-1-loop}, guaranteeing universality at one-loop. To ensure
universality at two-loops, there must be no hidden one-loop running couplings.
Suitable tuning of the seed action is sufficient to remove all but one~\cite{YM-2-loop}.

Referring back to equation~\ref{eq:SuperGaugeField}, we see that, in addition to our
physical gauge field $A^1_\mu$, we also have the unphysical gauge field $A^2_\mu$.
This carries its own coupling $g_2$, which renormalizes in a different way from $g$.
To obtain the universal value for $\beta_2$, we must isolate the effects of
$g_2$, within the ERG. Then, at the end of a computation, we will tune $g_2$ to zero.

\subsection{The New Flow Equation}

To isolate the effects of $g^2$, we modify the flow equation to allow it to 
distinguish between $A^1$ and $A^2$. This must be done 
 whilst, crucially, satisfying no-$\SF^0$ symmetry.
Introducing the new covariantized kernel $\{\dot{\Delta}^{\SF \SF}_{\sigma}\}$,
the flow equation receives the following modifications:
\begin{eqnarray*}
a_0[S,\Sigma_g] \rightarrow a_0[S,\Sigma_g] & + &
\frac{1}{16N} \fder{S}{\SF_\mu} \{\dot{\Delta}^{\SF \SF}_\sigma\}
\left[
\left\{\SH, \fder{\Sigma_g}{\SF_\mu} \right\} \str{\SH} - 2 \SH \str \left(\SH \fder{\Sigma_g}{\SF_\mu} \right)
\right] \\
& + & \frac{1}{16N} \left[
\left\{\SH, \fder{S}{\SF_\mu} \right\} \str{\SH} - 2 \SH \str \left(\SH \fder{S}{\SF_\mu} \right)
\right] \{\dot{\Delta}^{\SF \SF}_\sigma\} \fder{\Sigma_g}{\SF_\mu} \\
a_1[\Sigma_g] \rightarrow a_1[\Sigma_g] & + &
\frac{1}{16N} \fder{}{\SF_\mu} \{\dot{\Delta}^{\SF \SF}_\sigma\}
\left\{
\SH,  \fder{\Sigma_g}{\SF_\mu} \str \SH - \one \str \left(\SH \fder{\Sigma_g}{\SF_\mu} \right)
\right\} \\
& + & \frac{1}{16N}
 \left\{
\SH,  \fder{}{\SF_\mu} \str \SH - \one \str \left(\SH \fder{}{\SF_\mu} \right)
\{\dot{\Delta}^{\SF \SF}_\sigma\} 
\right\} \fder{\Sigma_g}{\SF_\mu}.
\end{eqnarray*}
Taking account of the running of $g_2$, the weak coupling flow equations are modified,
and the two-point tree level seed action vertices change; hence we must recompute the zero-point wines~\cite{YM-2-loop}.

\subsection{Computing $\beta$-function Coefficients}

With the formalism now in place, we can
calculate $\beta$-function coefficients. By computing the flow of
$S^{\ A^1 A^1}_{n \, \mu \ \nu}(p)$ and utilising the renormalization condition,
we obtain an algebraic expression for $\beta_n$, whose value can be extracted via
iterated use of the new flow equations.

Despite the changes to the flow equations, the diagrammatic methods of ref.~\cite{YM-1-loop}
can still be employed. Indeed in ref.~\cite{YM-2-loop} they have been greatly enhanced,
facilitating the computation of $\beta_2$. As anticipated, we find that, in the $g_2 \rightarrow 0$
limit, we regain the expected, universal coefficient demonstrating, beyond
all reasonable doubt, the consistency of the approach.

\paragraph{Acknowledgements} I would like to thank the organisers for a wonderful
conference, and the Southampton
University Development trust for financial support.

\end{document}